\documentclass[smallextended]{svjour2}                    % onecolumn
\smartqed  % flush right qed marks, e.g. at end of proof
\usepackage{amsmath,amsfonts}
\usepackage{graphicx}
\usepackage{comment}
\usepackage{amssymb,bm}
\usepackage[usenames,dvipsnames]{color}
\usepackage[version=3]{mhchem}
\usepackage{amssymb,bm}
\usepackage{amsmath,eucal,ulem,cancel}
\usepackage{graphics}

\journalname{Journal of Statistical Physics}
\begin{document}

\title{Detecting concentration changes with cooperative receptors
}
%\subtitle{Do you have a subtitle?\\ If so, write it here}

%\titlerunning{Short form of title}        % if too long for running head

\author{   Stefano Bo   \and  Antonio Celani   
         %etc.
}

%\authorrunning{Short form of author list} % if too long for running head

\institute{S. Bo \at {   Nordita\\
KTH Royal Institute of Technology and Stockholm University
Roslagstullsbacken 23, SE-106 91 Stockholm, Sweden 
             % Physics Department and INFN, University of Turin, via P. Giuria 1, I-10125 Turin, Italy
              \email{stefano.bo@nordita.org}      }
	\and
		A. Celani \at Quantitative Life Sciences, The Abdus Salam International Centre for Theoretical Physics (ICTP), Strada Costiera 11, I-34151 - Trieste, Italy \\
                   %  \\
%             \emph{Present address:} of F. Author  %  if needed
           }

\date{Received: date / Accepted: date}
% The correct dates will be entered by the editor

\maketitle

\begin{abstract}
 Cells constantly need to monitor the state of the environment to
 detect  changes and timely respond.
The detection of concentration changes of a ligand by a set of receptors can be cast as a problem of hypothesis testing, and the 
cell viewed as a Neyman-Pearson detector.
Within this framework, we investigate the role of receptor cooperativity in improving the cell's ability to detect changes.
We find that cooperativity decreases the probability of missing an occurred change. This becomes especially beneficial 
when difficult detections have to be made.
Concerning the influence of cooperativity on how fast a desired detection power is achieved, we find 
in general that there is an optimal value at finite levels of cooperation, even though easy discrimination tasks 
can be performed more rapidly by noncooperative receptors.
\\ 
%\PACS{  \and }
% \subclass{MSC code1 \and MSC code2 \and more}
\end{abstract}

\section{Introduction}
The ability to acquire, process and stock information is crucial for  living beings. 
% from single celled ones to higher organisms.
Already at the level of a single cell
 %, either as single organisms or in a multicellular assembly,
the external environment is chemically
sensed and  the acquired information is processed to prepare a suitable response.
All the steps involved in these processes are permeated with noise: the arrival of the external stimulus through diffusion is a 
stochastic process and so is the interaction with the receptors. The successive biochemical steps for the signal transduction usually
involve a limited amount of proteins so that fluctuations play a prominent role.
In other words, cells need to extract useful information from noisy inputs through noisy mechanisms, a seemingly daunting task.
Despite these hurdles, cells are able to sense the environment and make decisions with remarkable precision.
For a proper investigation of sensing it is fundamental to precisely characterize it and to assess its performance.
The first step towards a quantitative understanding of this issue was taken by Berg and Purcell
in their seminal paper in 1977~\cite{berg_purcell} addressing the question of how precise can a cell be in determining the concentration
of an external ligand.
Several studies have followed their line of reasoning
and considered signal to noise ratios as a measure of the quality of sensing
(see e.g. {  \cite{aquino11,aquino14,bialek05,bialek08,endres09,govern12,hu10,kaizu14,mora10,mugler13,rappel08,sevier14,skoge11,skoge13,sun14,tanase06}}). 
A complementary approach within the framework of information theory has been taken, investigating the ability of 
signaling networks to acquire information about the environment and transmit it  downstream 
(see Ref. \cite{bialek} for a general discussion, \cite{bowsher14,levchenko14} for  recent reviews
and 
\cite{bowsher12,cheong11,dubuis13,mancini13,martins11,mugler13,rieckh14,selimkhanov14,tkacik08,tkacik11,tkacik12,tostevin09,uda13,voliotis14,walczak10,ziv07}
for a representative though necessarily incomplete list of recent contributions).
% or the information flow (non distinguerli?)
% in transcriptional regulation \cite{bowsher12,dubuis,rieckh14,ziv07,mancini13}.
% and cite the ones discussed in \cite{marzen13} e.g \cite{martins11,tkacik08,,tkcacic11,tkacik12,walczak10}.
Furthermore, the influence of receptor cooperativity has been studied (see e.g.~{  \cite{aquino11,bialek08,sevier14,skoge11,skoge13,sun14}}).
Quantifying the ability to sense, transmit signals and respond to the external environment 
enables a quantitative comparison with its costs  
\cite{barato13pre,barato14,barato15,govern14pnas,govern14prl,hartich15,lan12,lang14,mancini15,mehta,murugan12,ouldridge15,qian05,qian06,sartori_pigolotti,sartori14,skoge13,tu08}.
This contributes to the understanding of the trade-offs which, under evolutionary pressure, may have shaped
the signaling and transcriptional strategies which are presently observed.

Here we focus on a different task: rather than precisely inferring the value of an external concentration, cells have to efficiently detect 
changes from a reference level.
Such systems include for instance the ones involved in enforcing homeostasis which must monitor deviations from the physiological
conditions, the early stages of the immune response with the detection of an antigen and more generally
signaling pathways downstream of receptors that undergo adaptation to the external stimuli.
The problem of detecting a change can be cast in terms of hypothesis testing.
Namely, one tests the hypothesis that 
a change has occurred {\it vs} the null one.
This problem has been thoroughly studied in Ref.~\cite{siggia13}, where it has been shown how
the occupation history of a single receptor can be used to perform a sequential probability ratio test.
In such dynamic formulation they derived how quickly a decision between two hypotheses can be made given an error threshold.
To carry out a statistical test exploiting the history of a receptor, cells need to devise molecular strategies
to record it and analyze it.  Refs. \cite{kobayashi10,siggia13} suggested some possible mechanisms
to encode statistical analysis in the level of some readout molecules which can be used to make decisions.
At variance with this approach, which focuses on sequential testing and considers a single receptor, here we study the case of parallel
testing through the instantaneous occupation state of a large pool of independent receptors.
This static approach no longer requires the presence of additional molecular layers as the receptors state at a given time can be used
directly as a readout. As a shortcoming, restricting to the instantaneous receptors occupation provides a less powerful test than the
 one exploiting the full receptor history.
 
We then view the  cell  as a  Neyman-Pearson detector 
which compares the likelihood of the observed receptor occupancy distribution
under the two hypotheses: the environment has not changed {\it vs} a change occurred. 
Within this framework we quantify the detection performance in terms of the
probability of missing the detection of a change.

%provides a more powerful test  
% than .
% Since the receptors state can be seen as a function (projection) of the full history,
% considering the Kullback-Leibler divergences, such intuition is formalized by the data processing inequality \cite{kullback}.
% The inequality states that the Kullback-Leibler divergence of a function of the stochastic variable 
% cannot exceed the one of the original one implying
% \begin{equation}
%  D_{KL}({\cal P}^{c}_t||{\cal P}_t^{c'})\ge D_{KL}(p_{eq}^c||p_{t}^{c'})\;.
% \end{equation}
% where ${\cal P}^{c}_t$ denotes the likelihood of the observed receptor path under the external concentration $c$.

The specific question that we address is whether cooperativity between the binding sites of a given 
receptor can be beneficial for effectively detecting changes. We find that cooperativity indeed
 increases the detection sensitivity
making it the preferable mechanism when difficult detections have to be performed.
When considering the time needed to achieve a certain statistical power we 
unveil a trade-off which results in an optimal finite level of cooperativity
which depends on the required sensitivity and on the number of binding sites
present in each receptor. 
In summary, we find that easy detections are often achieved more rapidly by receptors with low
levels of cooperativity (or even {   noncooperative} binding sites) whereas for difficult discrimination tasks a high cooperativity
is mandatory.

% As we will recall in the following sections the performance of multiple likelihood ratio tests 
% crucially depends on the Kullback-Leibler divergence of the probabilities of the two hypotheses.
% We therefore choose the Kullback-Leibler divergence to characterize the performance of the detection system.
% We focus on the case in which a cell can make various independent measurements at the same time,
% e.g. by means of independent membrane receptors or clusters.
% As a specific study system we consider a cell sensing the concentration of an external ligand by 
% means of some surface receptors. 
% The  binding and unbinding history of the receptors can be used to discriminate between the two hypotheses.
% 
%  We then address how the detection performance is affected by the level of detail about the receptors history available 
% for the discrimination. 
%  Namely, we compare the Kullback-Leibler divergence of the receptors trajectories to the one of
%  a projected observable more directly accessible: %not the full paths but
%   the instantaneous occupation state of the receptors. 
%   We provide as examples the case of independent and cooperative receptors highlighting their specific features.
%   In section \ref{sec:neyman} we recall the main concepts of Neyman-Pearson hypothesis testing and its relation with the
%   Kullback-Leibler divergence. We then consider the specific case of cooperative receptors described by Pauling model
%   in section \ref{sec:pauling}.

\section{Neyman-Pearson hypothesis testing for concentration discrimination}\label{sec:neyman}
\begin{figure}
\includegraphics[width=\columnwidth]{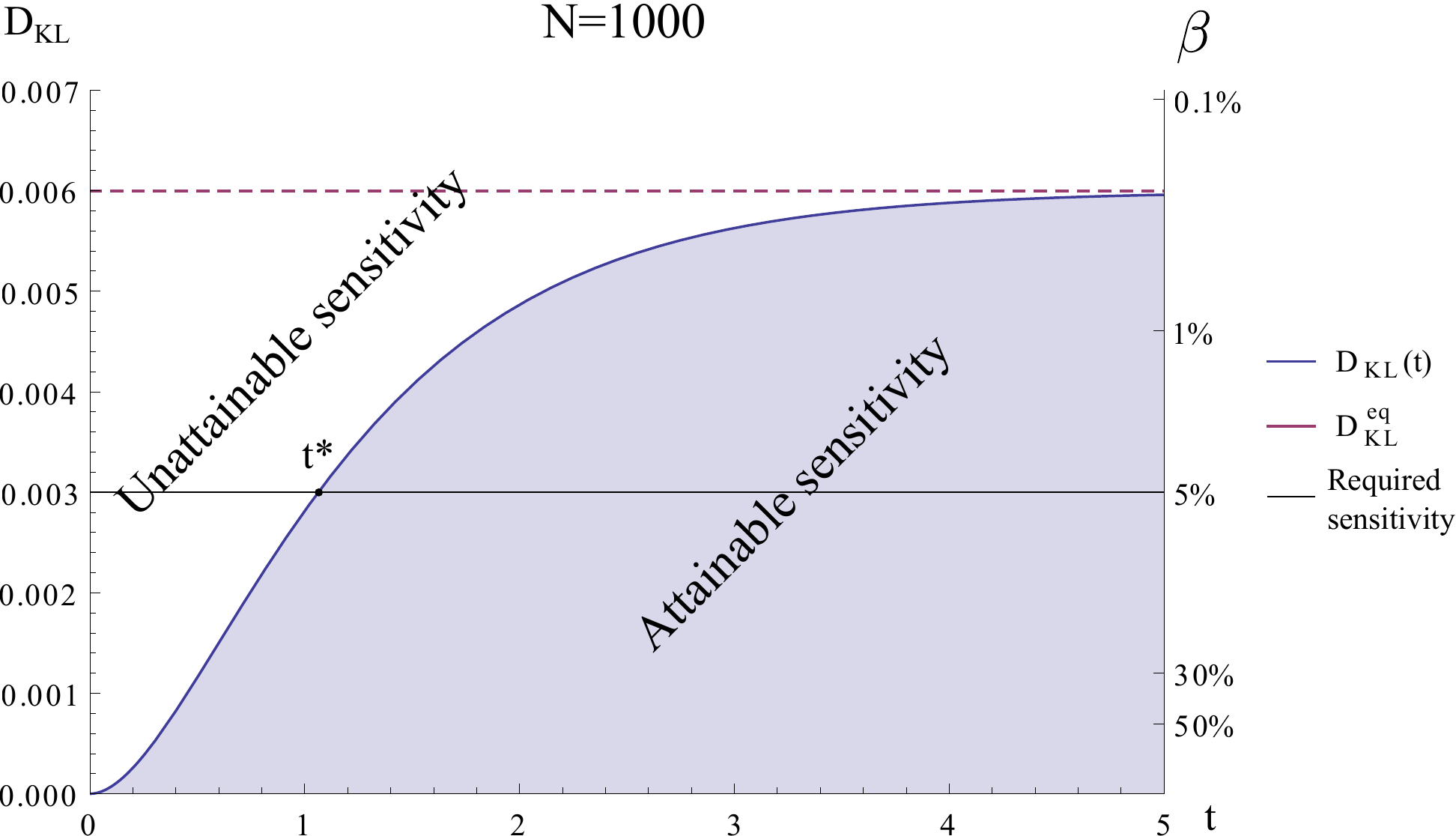}
\caption{Time evolution of the Kullback Leibler divergence $D_{KL}(p_{eq}^{c_{1/2}}||p_t^{c'})$ 
and of the associated miss probability $\beta$ for $N=1000$ receptors.
The level of the divergence sets the detection sensitivity. The shaded blue area refers to levels of sensitivity
which can be reached by the system.
As discussed in the text, the concentration change cannot be
immediately detected as the initial divergence is zero. As time goes by, the discrimination power increases 
until an equilibrium value is reached. If one is interested in a specific sensitivity (thin black line in the plot set at
a miss probability of $5\%$
corresponding to $D_{KL}=0.003$)
there is an associated time needed to cross the threshold $t^*$.
The plot is a numerical solution for the Pauling model described in section \ref{sec:pauling} for a receptor
with $L=3$ binding sites cooperating with a coupling of $J=1$. The concentration change to detect is
$c'=0.9*c_{1/2}$. 
Time is expressed in units of {   $\left(k_u\sqrt{c'/c_{1/2}}\right)^{-1}$} i.e. about the time needed for a (un-)binding event in the noncooperative
case. 
}\label{fig:time}
\end{figure}

%\subsection{Current receptors state}
To illustrate the main ideas of our approach let us start with a simple example.
Consider the case in which a cell has to determine whether the concentration of a given ligand has changed or not
by means of the occupation state of its receptors.
Such problem  can be addressed in terms of testing
 the null hypothesis  that no change occurred {\it vs} the alternative one that the concentration
 has changed.
 For the sake of simplicity let us restrict  to the specific case in which 
 the concentration has been constant for a long time and the 
 change is an instantaneous switch from a value $c$ to $c'$ taking place at time $t=0$.
  \begin{figure}
\includegraphics[width=\columnwidth]{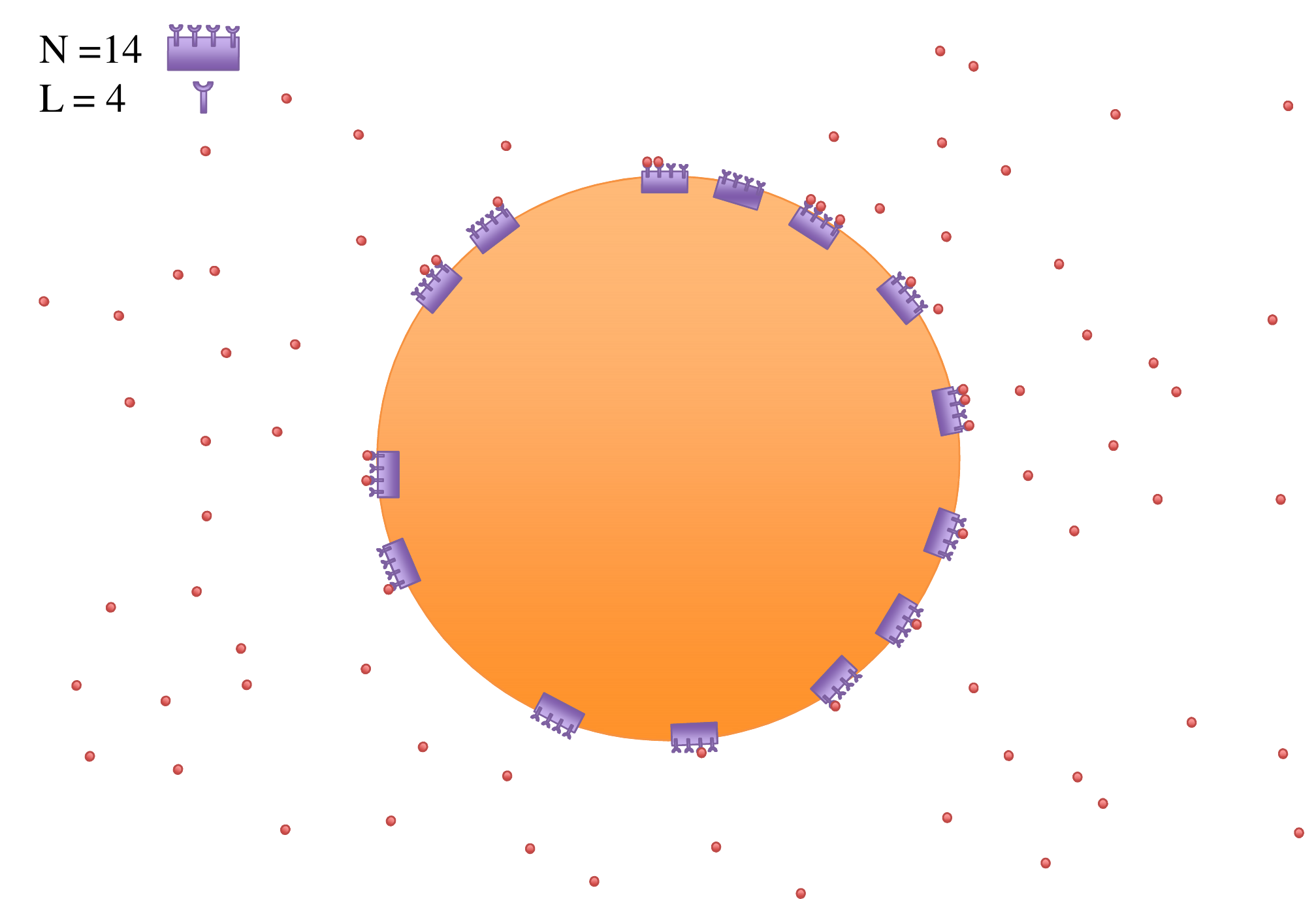}
\caption{Schematic view of a cell and its sensing components. The cartoon depicts a case with $N=14$ independent 
receptors each equipped with $L=4$ binding sites.
}\label{fig:cell}
\end{figure}
  If a cell is equipped with  $N \gg 1$ independent receptors  each with $L$ different binding sites (see figure~\ref{fig:cell}), 
  the state of the system at a given time is specified by 
% $\bm{X}=X_1,X_2,\ldots X_N$ where each $X_i=l_i=0,1,\ldots L$ denotes the number of occupied
% binding sites in receptor $i$. Since the receptors are independent we have that $\bm{X}$ is a collection of 
${\cal L}=l_1,l_2,\ldots l_N$ where each $l_n=0,1,\ldots L$ denotes the number of occupied
binding sites in receptor $n$. Since the receptors are independent we have that $\cal{L}$ is a collection of 
$N$ independent identically distributed variables drawn from the probability of the occupation number 
of a single receptor 
% at time $t$ evolving under a concentration $c$:
$p_t(l)$. Inferring if the concentration of the external ligand
has changed corresponds to choosing between two hypothesis
\begin{eqnarray}
 H_0:\, p_t(l)=p_0=p_{eq}^c(l)\\
 H_1:\, p_t(l)=p_1=p_{t}^{c'}(l)\nonumber
\end{eqnarray}
where $p_{eq}^c(l)$ is the equilibrium probability of a receptor having $l$ occupied binding sites when the concentration is $c$ and
$p_{t}^{c'}(l)$ is the time dependent one for a system that started in equilibrium with $c$  but for which the concentration switched to $c'$ at $t=0$.  
Neyman-Pearson lemma ensures that for a given significance $\alpha$ (probability of mistakenly rejecting the null hypothesis when it is 
true, type I error) the most powerful test (i. e. minimizing the probability $\beta$ of discarding the alternative hypothesis when it is
the correct one, type II error)
 is the likelihood ratio \cite{cover}. 
 Such test implies that for a single receptor the null hypothesis is rejected when $l$ is such that
\begin{equation}
 \Lambda(l)=\frac{p_{eq}^c(l)}{p_{t}^{c'}(l)}\le \eta
\end{equation}
 where $\eta$ determines the significance
 \begin{equation}
  \alpha=Prob(\Lambda\le \eta|H_0)=\sum_l p_{eq}^c(l)\Theta\left(\eta p_{t}^{c'}(l)-p_{eq}^c(l)\right)
 \end{equation}
where $\Theta$ is the Heaviside step function (i.e. $\Theta(x)=1$ if $x\ge 0$ and $\Theta(x)=0$ if $x< 0$).
 According to Stein's lemma, when 
 the number of independent receptors is large ($N \gg 1$) the probability of not detecting the change when it has occurred
 decreases exponentially as $\beta\sim e^{-N D_{KL}(p_0||p_1)}$ where the Kullback-Leibler divergence is defined as
 \begin{equation}
  D_{KL}(p_0||p_1)=\sum_l p_0(l)\log\frac{p_0(l)}{p_1(l)}
 \end{equation}
 which for this specific example reads:
 \begin{equation}\label{eq:kl_gen_proj}
  D_{KL}(p_{eq}^c||p_{t}^{c'})=\sum_l p_{eq}^c (l)\log\frac{p_{eq}^c(l)}{p_{t}^{c'} (l)}\;.
 \end{equation}
%Let us now focus on the projected Kullback-Leibler divergence of the receptor state (\ref{eq:kl_gen_proj}) and investigate its temporal behavior.
The divergence, and consequently the test sensitivity, changes with time (see fig.~\ref{fig:time}).
 At the very beginning the two distributions are equal and discrimination is therefore impossible.
%Since the Kullback-Leibler divergence is never negative this means that it displays a minimum at ti
For short times the likelihoods of the two hypotheses are still very similar and %in analogy to what discussed for small concentration changes 
 the Kullback-Leibler divergence grows quadratically in time (due to probability conservation there is no linear contribution).
 For long times the probability of the alternate hypothesis approaches the equilibrium value $\lim_{t\to\infty}p_{t}^{c'}=p_{eq}^{c'}$
 and the divergence saturates at the value $D_{KL}(p_{eq}^c||p_{eq}^{c'})$.
 %If the time evolution of the occupation probability is Markovian
 The existence of a stationary distribution $p_{eq}^{c'}$ ensures that
 the Kullback-Leibler divergence grows monotonically in time.
 This means that the asymptotic value $D_{KL}(p_{eq}^c||p_{eq}^{c'})$ sets the maximum detection precision attainable
 for the given concentration change.
 For intermediate times,  
the divergence changes concavity (at least once) from positive to negative. 

\section{Pauling model of cooperative receptors}\label{sec:pauling}
Let us now take a closer look at the probability distribution of the occupation level of an individual receptor
and study its dependence on the external concentration.
%We  consider the problem of sensing a concentration change by means of $N$ independent receptors.
We consider each receptor to consist of $L$ binding sites and  its read-out to be given by its number of 
occupied binding sites $l=0,\,1\,\ldots L$.
To investigate the role of cooperativity we choose the Pauling model in which each binding site of the receptor interacts with all
the other ones \cite{phillips_book}.
A given binding site $i=1,\ldots L$, can be occupied ($\sigma_i=0,\;1$) with a probability that depends on the external concentration of the ligand,
on its binding energy and, through cooperativity, on the number of other sites of the receptor which are bound.
The system 
can be described in terms of an Ising model with 
 Hamiltonian:
 \begin{equation}\label{eq:H_pa}
 H=-h\sum_{i=1}^L \sigma_i-\frac{J}{2}\sum_{i\neq j}\sigma_i\sigma_{j}
\end{equation}
where, for the sake of simplicity, we have encoded in $h$ both the contribution of the binding energy and the one of the chemical
potential which is affected by the ligand concentration:
\begin{equation}\label{eq:h_def}
 h=\log\frac{c}{K_d}
\end{equation}
{   where $K_d$ is the dissociation constant for the noncooperative receptor and we have set } $k_B T$ to unity for the rest of the paper.
We
refer to \cite{phillips_book} 
for a detailed connection between the statistical mechanical and the chemical description
of the systems.
Notice that, in this system, cooperativity
is encoded in the fact that the probability of binding 
increases with the number of bound sites.
Such cooperativity grows with the number of interacting binding sites and with the coupling parameter $J$.
Setting $J=0$ corresponds to considering the noncooperative case of independent binding sites.
% At variance with the locally cooperative system we have that t
The energy of a configuration depends only on the 
occupation number $l=\sum_{i=1}^L \sigma_i$ giving
\begin{equation}
 H(l)=-hl-\frac{J}{2}l(l-1)
\end{equation}
and the number of configurations with the same occupation number is simply given by the binomial coefficient ${L \choose l}$.
The equilibrium occupation probability then reads
\begin{equation}\label{eq:pa_eq}
 p_{eq}(l)=\frac{{L \choose l}\exp\left[ hl+\frac{J}{2}l(l-1)\right]}
 {\sum_{l=0}^L {L \choose l}\exp\left[ hl+\frac{J}{2}l(l-1)\right]}
\end{equation}
and the average number of bound sites can be directly computed as
\begin{equation}\label{eq:ave_l}
 \langle l \rangle=\sum_{l=0}^L l  p_{eq}(l)\;.
\end{equation}
The cooperativity of the system allows to have sharp changes in the occupation probability as the external concentration is varied.
The system is then said to be ultrasensitive around the value of the external field for which half of the binding 
sites are bound on average $h_{1/2}$, which we refer to as the transition point.
Such value depends on the number of total binding sites $L$ and the coupling $J$
\begin{equation}
 h_{1/2}=-\frac{J}{2}(L-1)
\end{equation}
and ensures that $p_{eq}(l)=p_{eq}(L-l)$.
We note that for highly cooperative systems the equilibrium distribution at $h=h_{1/2}$ is concentrated at the 
extremes $l=0$ and $l=L$.
The transition point occurs at different values of concentration as the number of binding sites or the coupling are changed
\begin{equation}\label{eq:c}
 c_{1/2}(J,L)=K_de^{- \frac{J}{2}(L-1)}\;.
\end{equation}
 To proceed to a meaningful comparison between
different cooperativities we need to consider systems that have to detect changes of the same relative amplitude $c/c_{1/2}$.
In general, cooperative receptors with  different numbers of binding sites or cooperative strengths can set their transition 
point at the same concentration by tuning their dissociation constant. This is a simple mechanism that implements sensorial adaptation.
% constant depending on the total number of binding sites $L$ and the 
% coupling $J$. 
% We are concerned with systems monitoring deviations from the reference condition $c_{1/2}$ which is set by the specific biological 
% problem at hand.
% To proceed to a comparison between systems that differ in the number of binding sites but display a transition 
% point for the same concentration we let the dissociation constant depend on the total number of binding sites $L$ and the 
% coupling $J$. 
Indeed, by letting
% \begin{equation}\label{eq:kd_scale}
 $K_d(L,J)=K_d^{(0)}e^{ \frac{J}{2}(L-1)}$
% \end{equation}
the transition point  is reached for the same concentration $c_{1/2}=K_d^{(0)}$.
This means that in order to maintain ultra-sensitivity around a given concentration
the receptor has
to compensate for a higher cooperativity by evolving towards
a higher dissociation constant (lower affinity with ligand).
% The relevant concentration is set by the biological problem at hand:
% physiological level of some molecule
% Indeed, some systems observed in nature, such as the bacterial chemotactic receptors
% adapt so that  as reported in \cite{keymer06}.
\subsection{Cooperativity and discrimination}
\begin{figure}
\includegraphics[width=\columnwidth]{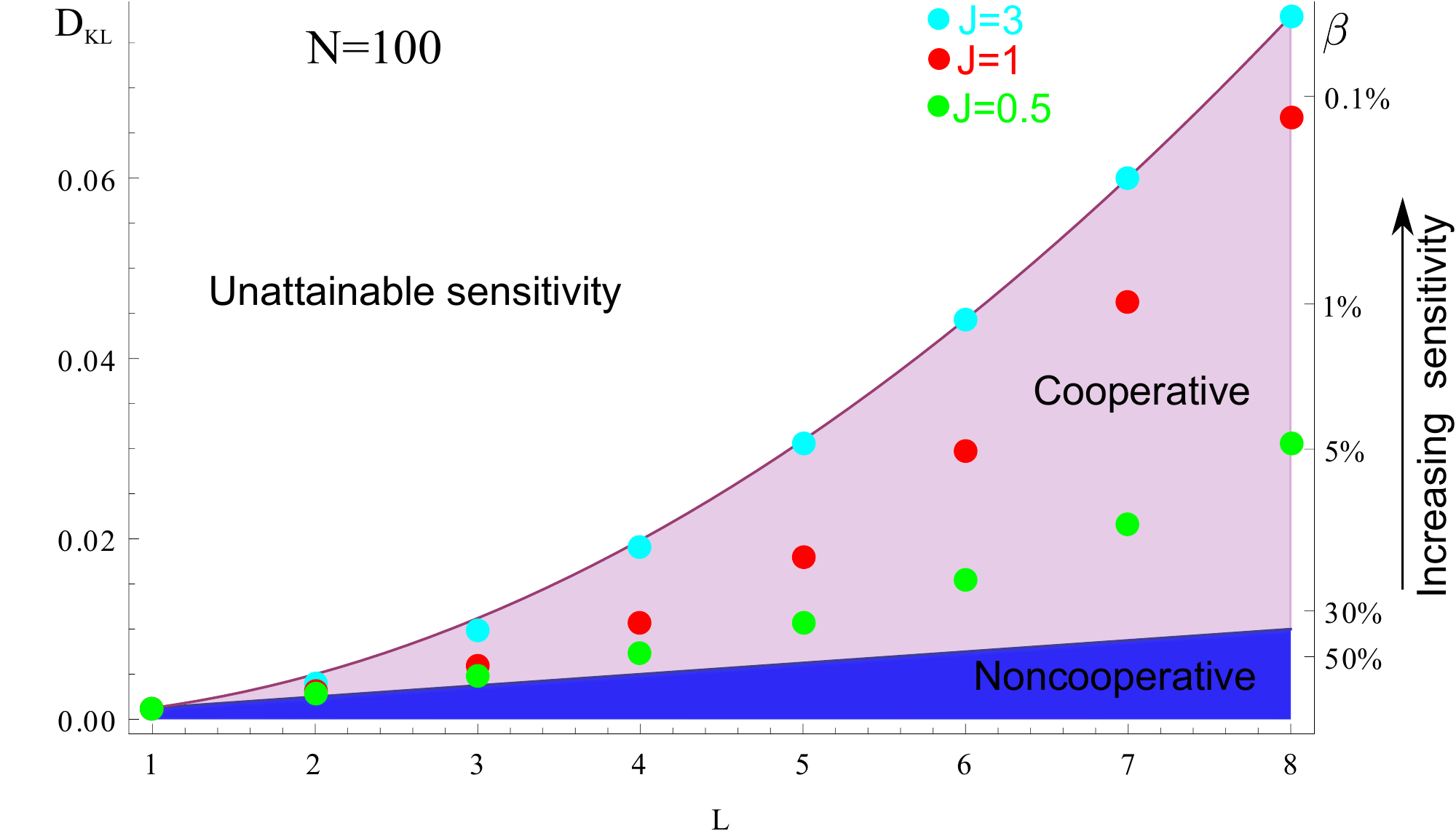}
\caption{Long-time asymptotic Kullback Leibler divergence and miss probability for $N=100$ receptors as a function of binding sites numbers for
% $h=h_{1/2}$ and $h'=h_{1/2}-0.1$. 
$c'=0.9*c_{1/2}$.
The full purple 
line is the divergence in the limit of infinite coupling ($J\to \infty$) discussed in eq.~(\ref{eq:pa_largeJ}) and the full blue one is
the noncooperative case ($J=0$) discussed in eq.~(\ref{eq:pa_ind}).
The shaded blue area depicts the values of precision %the Kullback-Leibler divergence 
reachable with {   noncooperative} binding sites,
the purple shaded one the values achievable by cooperative receptors. The white area above the purple line refers to 
 precisions
 %of Kullback-Leibler divergence
 which cannot be reached for the given number of binding sites. 
The cyan points are numerically evaluated results for $J=3$ already approaching the infinite coupling limit.
The red  points refer to the  case with $J=1$ and the green ones to $J=0.5$.
% As expected, cooperativity grows with the number of interacting binding sites and consequently the Kullback-Leibler
% divergence grows faster than the linearly increasing one of the independent case.
}\label{fig:pa_eq}
\end{figure}

As discussed above, around the transition point, the distribution of the occupation number of the receptor
exhibits a stronger dependence on  concentration as cooperativity
is increased.
Such sharper dependence %on concentration around the transition point
makes it easier to tell 
if the concentration has changed from the reference value (set at the transition point).
Intuitively, then, cooperativity improves the discrimination ability of the receptor.
To check this intuition let us consider the behavior of
the Kullback-Leibler distance between the equilibrium probability
at the transition point and the one of a different concentration and its dependence
on the degree of cooperativity (see figure~\ref{fig:pa_eq}).
We start by considering the asymptotic value of the divergence which corresponds to the largest
possible discrimination.
Due to Stein's lemma, the Kullback-Leibler divergence 
describes how the probability of missing a change in concentration decreases with the 
number of independent measurements $N$ in the limit $N\to \infty$. 
For instance to achieve a miss probability $\beta \simeq 5\%$ we need 
a divergence between the two distributions of 
\begin{equation}
 D_{KL}(p^{c_{1/2}}_{eq}||p^{c'}_{eq})=-\frac{\log \beta}{N}\simeq \frac{3}{N}\;.
\end{equation}
For a given number of independent receptors $N$ the Kullback-Leibler divergence then sets the miss probability.

Let us start by considering the noncooperative case ($J=0$). The system then simplifies greatly and is just a collection of $L$ independent
two states processes with binomial distribution
\begin{equation}\label{eq:p_eq(c)}
 p_{eq}^{\mathit{nc}}(l)={L\choose l}\frac{\left(\frac{c}{c_{1/2}} \right)^l}
 {\left(\frac{c}{c_{1/2}}+1 \right)^{L}}\;.
%   p_{eq}^{c}(l)={L\choose l}\frac{\left(\frac{c}{K_d^{(0)}} \right)^l}
%  {\left(\frac{c}{K_d^{(0)}}+1 \right)^{L}}\;.
\end{equation}
The Kullback-Leibler divergence is then analytically
accessible and considering the two equilibrium distributions reads:
\begin{equation}\label{eq:pa_ind}
D_{KL}^{\mathit{nc}}(p^{c_{1/2}}_{eq}||p^{c'}_{eq})=
%L\left[\log\frac{K_d+c'}{K_d+c}+\frac{c}{K_d+c}\log\frac{c}{c'} \right]=
%L\log\left(\frac{\sqrt{\frac{K_d^{(0)}}{c'}}+\sqrt{\frac{c'}{K_d^{(0)}}}}{2}\right)=
L\log\left(\frac{\sqrt{\frac{c_{1/2}}{c'}}+\sqrt{\frac{c'}{c_{1/2}}}}{2}\right)=
L\log\left[\cosh\left( \frac{ h'-h_{1/2}}{2}\right)\right]
\end{equation}
where
% we have made explicit use of the fact that for independent receptors
% $c_{1/2}=K_d$ and
we shall recall that $h_{1/2}=0$.
The most relevant feature is that the divergence grows linearly with the number of binding sites present in the receptor.

We expect cooperativity to increase the Kullback-Leibler divergence and consequently decrease the probability of not detecting a concentration
change. In the limit of large coupling $J\to \infty$  we can roughly estimate the occupation probability
by inspecting equation (\ref{eq:pa_eq}) around $h_{1/2}=-J(L-1)/2$ and observing that only states with $l=0$ and $l=L$ will have a finite probability:
\begin{eqnarray}
 \lim_{J\to\infty} p(l=0)\simeq \frac{1}{1+e^{  L(h'-h_{1/2})}}=\frac{1}{1+\left(\frac{c'}{c_{1/2}}\right)^L}\\ \nonumber
 \lim_{J\to\infty} p(l=L)\simeq \frac{e^{  L(h'-h_{1/2})}}{1+e^{  L(h'-h_{1/2})}}=
 \frac{\left(\frac{c'}{c_{1/2}}\right)^L}{1+\left(\frac{c'}{c_{1/2}}\right)^L}
\end{eqnarray}
giving
\begin{eqnarray}\label{eq:pa_largeJ}
 \lim_{J\to\infty} D_{KL}(p^{c_{1/2}}_{eq}||p^{c'}_{eq})&=&
%L\left[\log\frac{K_d+c'}{K_d+c}+\frac{c}{K_d+c}\log\frac{c}{c'} \right]=
\log\left(\frac{\left(\frac{c_{1/2}}{c'}\right)^{L/2}+\left(\frac{c'}{c_{1/2}}\right)^{L/2}}{2}\right)\\\nonumber
&=&\log\left[\cosh\left(\frac{  L (h'-h_{1/2})}{2}\right)\right]
\end{eqnarray}
which for $L>1$ is larger than its noncooperative equivalent.
Hence, for a fixed $L$, having a cooperative receptor allows to reach values of the Kullback-Leibler that cannot 
be approached by means of {   noncooperative} binding sites resulting in higher detection sensitivities.
However, the limit in eq.~(\ref{eq:pa_largeJ}) cannot be exceeded and this sets an upper bound on the performance of a cooperative receptor following
Pauling model.
% \begin{figure}
% \includegraphics[width=0.8\columnwidth]{kl_cont.pdf}
% \caption{Contour plot of Kullback Leibler divergence between equilibirum distributions as a function of binding sites numbers $L$ and coupling strength $J$ 
% for $\beta=1$, $h=h_{1/2}$ and $h'=h_{1/2}-0.1$. As expected (and shown in fig.~\ref{fig:pa_eq}) the divergence grows with the number of binding sites 
% and the coupling strength.
% }\label{fig:kl_cont}
% \end{figure}

% Considering a small concentration change, we see that
% the discrimination power of a receptor with cooperative binding sites increases more than linearly with $L$ as shown in fig.~\ref{fig:pa_eq}.
% The figure also shows how discrimination is enhanced by increasing the coupling between binding sites at finite $J$. 
% Fig.~\ref{fig:kl_cont} reports the same behavior by means of a contour plot.
\subsubsection{Small concentration change limit}
The hypotheses we are testing have probability distributions that differ because of the 
parameter $c$ and $c'$.
Obviously, when the two concentrations are equal the Kullback-Leibler divergence is zero and is at a minimum since it is 
non-negative by definition.
Then, for small concentration differences $\delta c/c\ll 1$, the divergence is quadratic
 in the concentration difference.
%  This can be understood by considering that the divergence has a minimum for $c=c'$ where 
%  it therefore has a vanishing derivative.
 It is known that  curvature of the divergence in $c=c'$
 is equal to the Fisher information of the probability distribution with respect to the 
 parameter $c$ which is defined as
 \begin{equation}
 {\cal I}(c)=\sum_{l=0}^L\left( \frac{\partial \log{p^c_{eq}(l)}}{\partial c}\right)^2 p^c_{eq}(l)\,.
\end{equation}
Then, by Taylor expansion, the  Kullback-Leibler for small concentration changes around 
%$c_{1/2}$
 {   $c$ reads 
\begin{equation}\label{eq:fisher}
\lim_{\delta c/c\to 0} D_{KL}(p^c_{eq}||p^{c+\delta c}_{eq})\simeq \frac{1}{2}\left(\delta c\right)^2 {\cal I}(c)\;.
\end{equation}
}
%For small concentration changes we can exploit the relation between the Fisher information and the Kullback-Leibler divergence.
Plugging relation (\ref{eq:h_def}) into the equilibrium probability (\ref{eq:pa_eq}) we can compute the derivative
of the probability with respect to concentration and obtain:
\begin{equation}
 \frac{\partial p^c_{eq}(l)}{\partial c}=\frac{l-\langle l \rangle}{c}p^c_{eq}(l)
\end{equation}
which means that the Fisher information is
$
 {\cal I}(c)=\frac{1}{c^2}var(l)
$
and the Kullback-Leibler for small concentration changes {   around the transition point $c_{1/2}$}
\begin{equation}
\lim_{\delta c/c\to 0} D_{KL}(p^{c_{1/2}}_{eq}||p^{c_{1/2}+\delta c}_{eq})\simeq \frac{1}{2}\left(\frac{\delta c}{c_{1/2}}\right)^2var(l)\;.
\end{equation}

It is interesting to compare the two extremes of large cooperativity and vanishing cooperativity.
Taking the large coupling limit $J\to\infty$ of the variance we see that
\begin{equation}
\lim_{J\to \infty}\lim_{\delta c/c\to 0}D_{KL}(p^{c_{1/2}}_{eq}||p^{c_{1/2}+\delta c}_{eq})\simeq 
\lim_{J\to \infty}\frac{1}{2}\left(\frac{\delta c}{c_{1/2}}\right)^2var(l)
\simeq \frac{1}{8}\left(\frac{\delta c}{
c_{1/2}}\right)^2L^2
\end{equation}
where two limits can be shown to commute by considering eq.~(\ref{eq:pa_largeJ}). 
For the noncooperative case one finds
{  
\begin{equation}
\lim_{\delta c/c\to 0}D^{\mathit nc}_{KL}(p^{c_{1/2}}_{eq}||p^{c_{1/2}+\delta c}_{eq})\simeq 
\frac{1}{8}\left(\frac{\delta c}{
c_{1/2}}\right)^2L\;.
\end{equation}}
The strongly cooperative case displays a quadratic dependence on the number of binding sites, as opposed to the linear one of
the noncooperative model. The miss probability of the test then decreases exponentially with $L^2N$ for the cooperative
case and simply as $LN$ for the noncooperative one.
{   
For finite couplings, 
the variance has to be computed directly from the distribution and reads:
\begin{eqnarray}
&&var(l)={\sum_{l=0}^L \left(l-\langle l\rangle\right)^2p_{eq}(l})\\\nonumber
&&=
\frac{ \sum_{l=0}^L l^2 {L \choose l}\exp\left[ hl+\frac{J}{2}l(l-1)\right]}
 {\sum_{l=0}^L {L \choose l}\exp\left[ hl+\frac{J}{2}l(l-1)\right]}
 -\left(   \frac{ \sum_{l=0}^L l {L \choose l}\exp\left[ hl+\frac{J}{2}l(l-1)\right]}
 {\sum_{l=0}^L {L \choose l}\exp\left[ hl+\frac{J}{2}l(l-1)\right]}   \right)^2
\end{eqnarray}
Such expression, due to the terms $\exp\left[ \frac{J}{2}l(l-1)\right]$  does not display a simple scaling in terms of $L$  for finite couplings.

\subsubsection{Connection between Hill coefficient and Kullback-Leibler divergence}
The Hill coefficient is a measure of the cooperativity of binding and can be defined (see e.g. Ref.~\cite{hill85}) as:

%\begin{equation}
%n_H=\frac{\partial \log\frac{ \langle l \rangle}{L- \langle l \rangle}}{\partial \log c}\biggr|_{c=c_{1/2}}\;.
%\end{equation}
%At the transition point $ \langle l \rangle=L- \langle l \rangle=L/2$ so that we can write

  \begin{figure}
\includegraphics[width=\columnwidth]{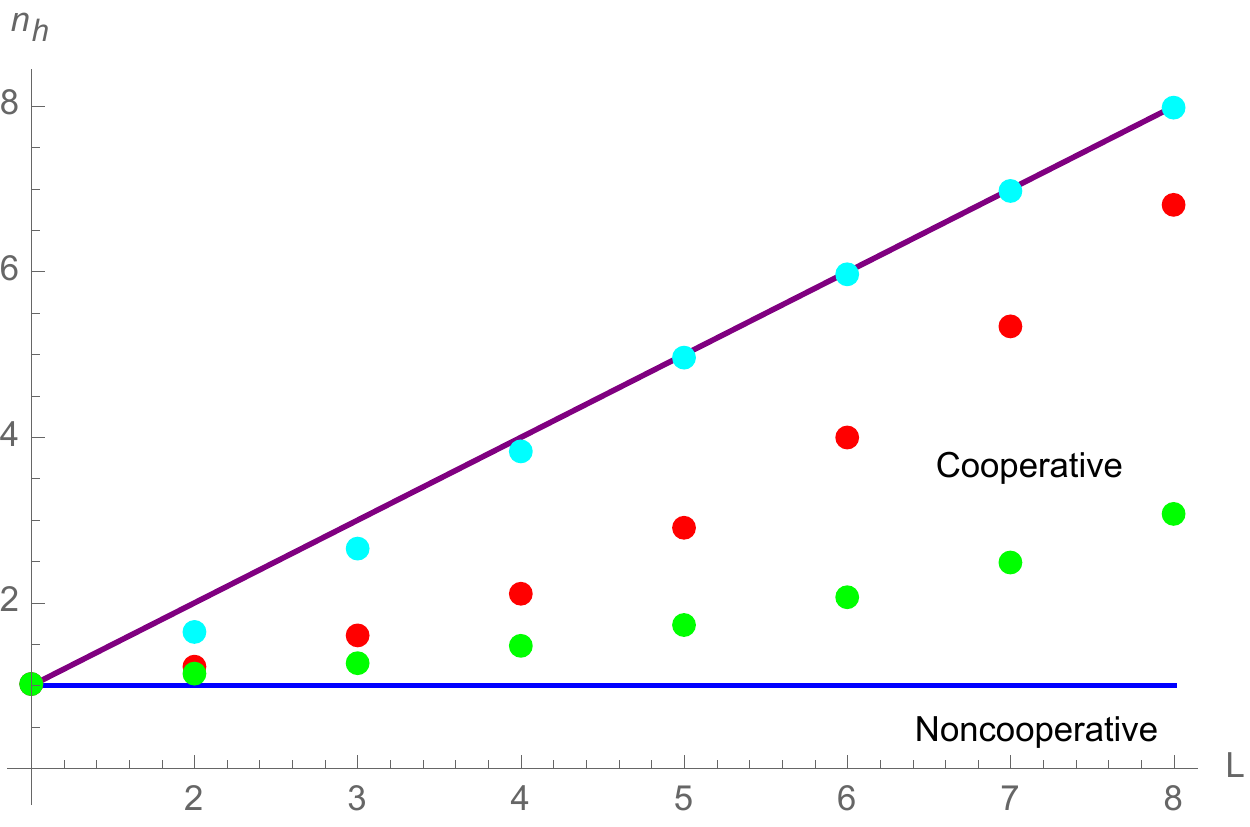}
\caption{Hill coefficient as a function of binding sites per receptor $L$ for different values of the coupling $J$. For the noncooperative case 
the Hill coefficient is $1$ whereas it is $L$ in the limit of strong cooperativity. The parameters and color code are the same as for
figure \ref{fig:pa_eq}.
}\label{fig:hill}
\end{figure}

\begin{equation}
n_H=\frac{4}{L}\frac{\partial \langle l \rangle}{\partial \log c}\biggr|_{c=c_{1/2}}=\frac{4}{L}\frac{\partial \langle l \rangle}{\partial h}\biggr|_{c=c_{1/2}}\;.
\end{equation}
Performing the derivative of the mean value with the explicit expressions given in eqs.~(\ref{eq:pa_eq}) and (\ref{eq:ave_l})
we see that\footnote{This is simply the fluctuation dissipation relation between the susceptibility and the magnetization variance
for ferromagnetic systems}
\begin{equation}
\frac{\partial \langle l \rangle}{\partial h}=var(l)\;.
\end{equation}
 We can then write 
\begin{equation}
n_H=\frac{4}{L }var(l)\;.
\end{equation}

As we have seen in eq.~(\ref{eq:fisher}), for small concentration changes, the Kullback-Leibler divergence is proportional to the Fisher information
which,
 for the system we are considering,  is itself proportional
to the variance of the distribution ${\cal I}=var(l)/c^2$. Hence, for small concentration changes around the transition point, we have that
\begin{equation}\label{eq:link}
\lim_{\delta c/c\to 0} D_{KL}(p^{c_{1/2}}_{eq}||p^{c_{1/2}+\delta c}_{eq})\simeq \frac{1}{2}\left(\frac{\delta c}{c}\right)^2var(l)=\frac{L}{8}\left(\frac{\delta c}{c}\right)^2 n_H
\end{equation}
linking the Kullback-Leibler divergence and the Hill coefficient. 
The Hill coefficient ranges from $1$ for noncooperative binding to a maximum of $L$ approached in the limit of infinite cooperativity (see fig. \ref{fig:hill}).
This offers some insight on the scaling behavior of the Kullback-Leibler divergence for finite couplings.
Indeed, for small concentration changes, the Kullback-Leibler divergence scales as $Ln_H$ where $n_H$ is
the Hill coefficient of the receptor. }
\subsection{How long does it take to detect a change?}
From figure~\ref{fig:time} we know that at the moment in which the concentration changes it is not possible yet to detect the change
as the distributions of the two hypotheses are still equal.
Detection becomes possible as the likelihood of the alternative hypothesis evolves in time.
We are interested in studying how long it takes before the distributions of the two hypotheses 
are different enough to allow a detection with a given sensitivity.
In order to do so, let us now consider the kinetic behavior of a receptor whose binding sites follow the Pauling model.
In general the occupation probability evolves according to a master equation
\begin{equation}
 \frac{dp(l)}{dt}=\sum_l k(l'\to l)p(l')-k(l\to l')p(l)
\end{equation}
where, for the system to reach the correct equilibrium, detailed balance must be satisfied by the rates:
\begin{equation}
 \frac{k(l\to l')}{k(l'\to l)}=\frac{p_{eq}(l')}{p_{eq}(l)}=e^{-\left(H(l')-H(l)\right)}{L \choose l'}/{L \choose l}
\end{equation}
which for the two allowed transitions implies
\begin{eqnarray}
 \frac{k(l\to l+1)}{k(l+1\to l)}=\frac{L-l}{l+1}e^{h+Jl}=\frac{(L-l)c}{(l+1)c_{1/2}(J,L)}e^{J\left(l-\frac{L-1}{2}\right)}\;.
\end{eqnarray}
There is freedom in how to choose the individual rates still satisfying detailed balance.
We consider the case in which the rates are exponential in the energy difference associated
with the transition
%:
% % \begin{eqnarray}
% %  k(l\to l+1)&=&a(L-l) e^{  g \left(h+Jl\right)}\\
% %  k(l\to l-1)&=&a\,l e^{  (g-1) \left(h+J(l-1)\right)}
% % \end{eqnarray}
% % where $a$ is an overall constant and $g$ can be used to set how the effect of cooperativity
% % is split between the binding and unbinding rate: i.e. if it 
% % acts by speeding up binding 
% % or slowing down unbinding when $l$ grows.
% % the case in which unbinding is unaffected by the other binding sites state and binding is sped up as the number
% % of bound sites increases giving:
% % \begin{equation}
% %  k_+(l)=(L-l)e^{  h} e^{  J l}k_u \qquad \qquad k_-(l)=l k_u
% % \end{equation}
% % where as before $h=\frac{1}{ }\log\frac{c}{K_d}$ and as usual $K_d=k_u/k_b$.
% % In general, even when the dependence on $L$ is not quadratic the slope with which the cooperative KL increases with $L$ is higher than 
% % the one for independent binding sites.
% \begin{eqnarray}
%  k(l\to l+1)&=&a(L-l) e^{\frac{1}{2} \left(h+Jl\right)}\\
%  k(l\to l-1)&=&a\,l e^{-\frac{1}{2}  \left(h+J(l-1)\right)}
% \end{eqnarray}
% where $a$ is an overall constant which, 
 and to make contact with the non-cooperative binding  we set 
% to 
% \begin{equation}
%  a(J,L)=k_u\sqrt{\frac{c}{c_{1/2}(J,L)}}
% \end{equation}
% so that 
them to
% can be expressed as
\begin{eqnarray}
 k(l\to l+1)&=&(L-l)k_u \frac{c}{c_{1/2}(J,L)} e^{\frac{  J}{2} \left(l-\frac{L-1}{2}\right)}\label{eq:k_p}\\
 k(l\to l-1)&=&l\,k_u  e^{-\frac{  J}{2} \left(l-\frac{L+1}{2}\right)}\label{eq:km}
\end{eqnarray}
where $k_u$ is the unbinding rate for {   noncooperative} binding sites.
We can then proceed to the numerical solution of the master equation and obtain the time evolution of the system.
We shall focus on the case in which the system starts in equilibrium with the concentration at the transition point $c_{1/2}$ and
the concentration is switched to a different value $c'$.
The most remarkable feature is that, for fixed $c'/c_{1/2}$, %$\frac{c'}{c_{1/2}}$,
as cooperativity is increased (both by having more binding sites or by a higher coupling $J$)
%when there are several total binding site 
the system slows down. The slowing down is more severe 
for small concentration changes.
% as shown in figure~\ref{fig:pa_slow}. 
This phenomenon is related to the critical slowing down (discussed also in Ref.~\cite{skoge11})
and is due to the fact that a system with strong interactions is ``locked'' into a configuration due to cooperativity and 
reacts more slowly to a change in the external field. \\
% Indeed, as discussed earlier, at strong cooperativities the probability is concentrated 
% in the two extreme cases of $l=0$ and $l=L$
% \begin{center}
% 
% 
% \begin{figure}
% \includegraphics[width=0.8\columnwidth]{slow_small.pdf}
% \includegraphics[width=0.8\columnwidth]{slow_large.pdf}
% \caption{Critical slowing down. Evolution of relative mean occupation $\langle l\rangle/L$ for different number of binding sites with $J=1$, $ =1$ as 
% the concentration is switched from $c_{1/2}=K_d$ to $c'$.
% The blue curve refers to the non cooperative case, the red one to the cooperative case with $L=2$, the purple one to $L=5$, the yellow line
% to $L=8$ and the green one to $L=10$. The top panel refers to the small concentration change $c'=0.9K_d$ and the bottom to the larger
% change $c'=K_d/e\simeq0.368K_d$. Notice how the case of large concentration difference displays a faster equilibration compared to the 
% small concentration one. Time is expressed in units of $a=k_u\sqrt{c'/c_{1/2}}$.
% }\label{fig:pa_slow}
% \end{figure}
%  
% \end{center}

\begin{figure}
\includegraphics[width=0.8\columnwidth]{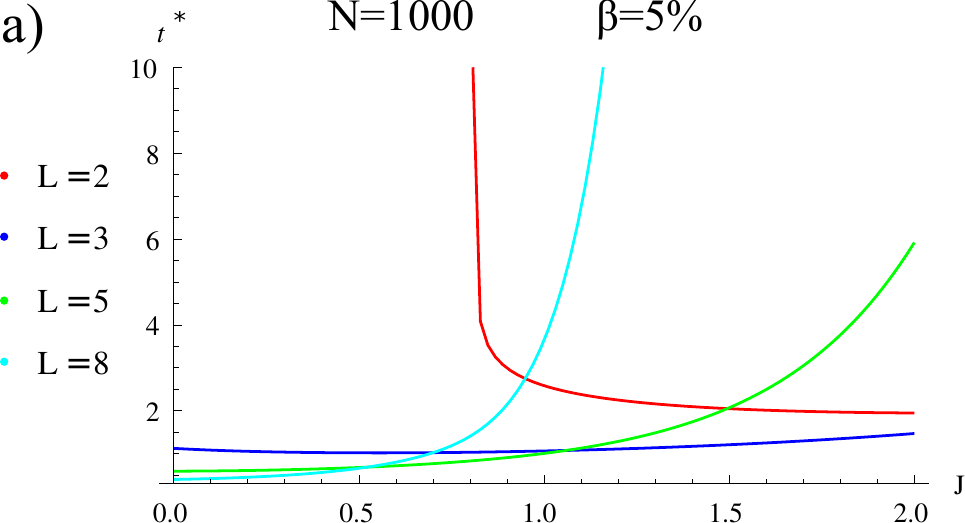}\\

\includegraphics[width=0.8\columnwidth]{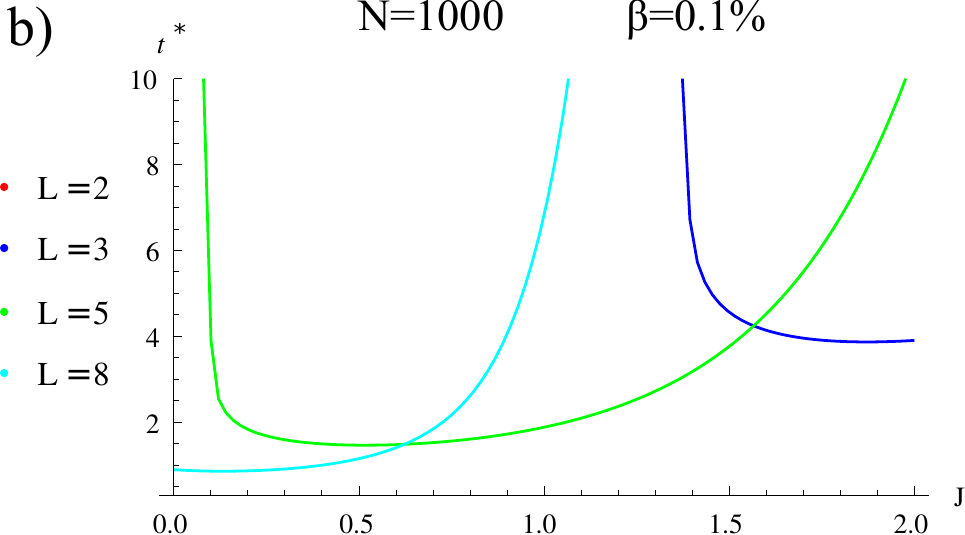}
\caption{Time needed to achieve a desired sensitivity in detecting a concentration change
to $c'=0.9*c_{1/2}$
for different numbers of binding sites and cooperativity.
This corresponds to time needed
by the Kullback Leibler divergence $D_{KL}(p_{eq}^{c_{1/2}}||p_t^{c'})$ to cross
a given  threshold. 
a): miss probability of $5\%$ which for $N=1000$ receptors corresponds to 
 $D_{KL}=0.003$. For $J<0.81$ the receptor with $L=2$ binding sites cannot reach
a divergence of $0.003$. 
For $L=3$ the shortest time is obtained at finite coupling whereas for $L=5\,,8$ for the noncooperative case. 
Notice the exponential slowing down observed for $L=8$.
b): miss probability of $0.1\%$  corresponding for  $N=1000$ receptors to  $D_{KL} \simeq0.0069$.
Now the required sensitivity is unattainable with $L=2$ binding sites. It requires a coupling $J>1.37$ for $L=3$
and a weak $J>0.1$ for $L=5$.
For the shown number of binding sites the desired precision is fastest reached for finite level of cooperativity.  
% Time is expressed in units of $a=k_u\sqrt{c'/c_{1/2}}$. 
Time is expressed in units of {   $\left(k_u\sqrt{c'/c_{1/2}}\right)^{-1}$ }i.e. about the time needed for a (un-)binding event in the noncooperative
case.
}\label{fig:time_det}
\end{figure}

Let us now investigate how this impacts on the time required by the pool of receptors to reach a certain detection sensitivity.
We need to consider the dynamic behavior of the Kullback-Leibler divergence during equilibration to the new
concentration $c'$ and how rapidly it manages to reach the threshold associated with the desired sensitivity as sketched in
figure~\ref{fig:time}.
As we have discussed in the previous section and  shown in figure~\ref{fig:pa_eq},
% and \ref{fig:kl_cont} 
the divergence between the equilibrium distribution in $c_{1/2}$ and
in $c'$ is higher for cooperative receptors consequently allowing to reach levels of sensitivity which are unfeasible for {   noncooperative}
binding sites. % more sensitive detections. % and this feature gets more marked with the increase of the number of binding sites.
However, a larger cooperativity slows down the dynamics. 
When considering how fast  a certain level is reached we are then facing a trade-off as shown in figure~\ref{fig:time_det}.
% and for short times a receptor with independent binding sites can have a higher Kullback-Leibler
% divergence than a cooperative one as shown in figure~\ref{fig:pa_kin}.
% \begin{center}
% \begin{figure}
% \includegraphics[width=0.8\columnwidth]{pauling.pdf}
% \caption{Kullback Leibler divergence as a function of time for different numbers of binding sites and cooperativity.
% The blue and the cyan curve refers to the non cooperative case respectively with $L=8,\,10$.
% The yellow  and the green line are for a cooperative receptor with $J=1$ respectively for
%  $L=8,\,10$ binding sites. The initial concentration is $c=c_{1/2}=K_d^{(0)}$ and is changed to  $c'=0.9K_d{(0)}$ i.e. a small concentration change.
% The cooperative cases reach a higher equilibrium value but are slower than the independent ones and are therefore lower at short times. 
% Time is expressed in units of $a=k_u\sqrt{c'/c_{1/2}}$
% }\label{fig:pa_kin}
% \end{figure}
% % \end{center}
% To better investigate the issue let us focus on the time needed to reach a certain threshold of the Kullback-Leibler divergence.
% As mentioned before, this corresponds to determining how long it takes before the pool of receptors is able to achieve a fixed
% statistical power.
% For a given number of binding sites, cooperativity allows to reach levels of sensitivity which are unfeasible for independent
% binding sites. 

To obtain some insight let us study a specific example.
Consider for instance the case in which we have $N=1000$ receptors, the concentration decreases of about $10\%$ from its transition value:
$c'=0.9*c_{1/2}$ and we want to achieve a miss probability of  $\beta\simeq 5\%$. This sets the Kullback Leibler distance between the equilibrium 
receptor distribution at the transition point and the evolving one at
$D_{KL}(p_{eq}^{c_{1/2}}||p_t^{c'})=0.003$. From eq.~(\ref{eq:pa_ind}) we know that such sensitivity cannot be achieved by
a receptor with $2$  binding sites unless
they operate in a cooperative fashion. 
% Figure~\ref{fig:time_det} shows the time needed to reach a Kullback-Leibler divergence of $0.003$ for different number of 
% binding sites and cooperative strengths.
%
% \begin{figure}
% \includegraphics[width=0.8\columnwidth]{time_cont.pdf}
% \caption{Time needed by the Kullback Leibler divergence $D_{KL}(p_{eq}^{c_{1/2}}||p_t^{c'})$ to cross
% the threshold $0.003$
%  for different numbers of binding sites and cooperativity with $c'=0.9*c_{1/2}$.
% The case with $2$ binding sites can reach the required Kullback-Leibler only in the cooperative case with $J>0.81$. 
% For the shown range, the fastest way to reach the desired sensitivity is to use receptors with $L=8$ independent binding sites.
% The slowest cases are for  $L=2$ when the threshold is barely reachable ($J\simeq0.81$) and for $L=8$ and $J\simeq1$ when
% the slowing sown of the dynamics dominates the benefit of a higher equilibrium value.
% Time is expressed in units of $a=k_u\sqrt{c'/c_{1/2}}$. 
% }\label{fig:time_cont}
% \end{figure}
Then, for small couplings, a receptor with $2$ binding sites will never reach the threshold. As the coupling is increased 
the threshold becomes attainable and it progressively takes shorter times to approach it until {   an} optimal value is reached.
When several binding sites can be employed the situation is quite the opposite as the threshold is largely exceeded
also in the noncooperative case so that the main effect of cooperativity is the exponential slowing down of the dynamics.
In such case the required sensitivity is reached earlier by {   noncooperative} binding sites.
For intermediate scenarios, as for example with $3$ binding sites, the noncooperative case is sufficient to
cross the threshold but its asymptotic value is not much larger. Then, a weak cooperativity is slightly faster as a result
of the trade-off between a higher equilibrium limit and slower overall dynamics.
The representative cases of $L=2,\,3,\, 5\,,8$ are plotted in figure~\ref{fig:time_det}.
Let us study how requiring a higher sensitivity with a miss probability as low as $\beta\simeq 0.1\%$ 
% we need to achieve
% a divergence $D_{KL}(p_{eq}^{c_{1/2}}||p_t^{c'})=\log(1000)/1000\simeq0.0069$.
impacts the trade-off.
We first notice that such low miss probability is not achievable if each receptor has only 2 binding sites since  $N=1000$ noncooperative such receptors
can at most reach a miss probability of $\beta\simeq 0.7\%$.
% which can at most have a Kullback-Leibler of
% $D_{KL}(p_{eq}^{c_{1/2}}||p_{eq}^{c'}) \simeq0.005$.
For $L=3,\,5$ the required sensitivity can be attained only by cooperative receptors and 
it is most rapidly reached for intermediate coupling values.
Finally, employing $8$ binding sites cooperativity is no longer necessary to reach the threshold.
However, a small coupling ($J\simeq 0.14$)  
allows to cross the desired value at slightly shorter times. 

In general, considering how cooperativity affects the time need for a sensitive detection we can identify three main classes 
which are determined by the level of desired sensitivity
and the number of available binding sites.
Namely, what matters is how large  the long-time asymptotic discrimination power of the noncooperative receptors
(eq. \ref{eq:pa_ind}) is compared to the one we want to obtain.
If the required sensitivity is easily achievable
by  noncooperative receptors (threshold much lower than the
equilibrium  value for the noncooperative receptors), cooperativity is detrimental for a rapid decision (see e.g. the case of $L=8$ binding
sites and $\beta=5\%$ depicted in cyan in figure~\ref{fig:time_det}.a).
For more sensitive detections, which are barely attainable by {   noncooperative} binding sites, a small degree of cooperativity
can slightly
speed up the system as shown for instance for $L=3$ and $\beta=5\%$ in blue in figure~\ref{fig:time_det}.a.
For hard detections (desired sensitivity higher than the one provided by noncooperative receptors) the system must be cooperative.
There is a large but finite coupling minimizing the time needed to achieved the threshold as reported for  $L=3$ and $L=5$ for
a miss probability of $\beta=0.1\%$ in figure~\ref{fig:time_det}.b (respectively blue and  green line).\\
{

So far,  we have considered the time needed to reach a certain level of precision as a function of the coupling intensity $J$ for the
case in which the reference concentration is set at the transition point $c=c_{1/2}$.
 \begin{figure}[h]
\includegraphics[width=0.45\columnwidth%, trim=1cm 2cm 3cm 4cm
]{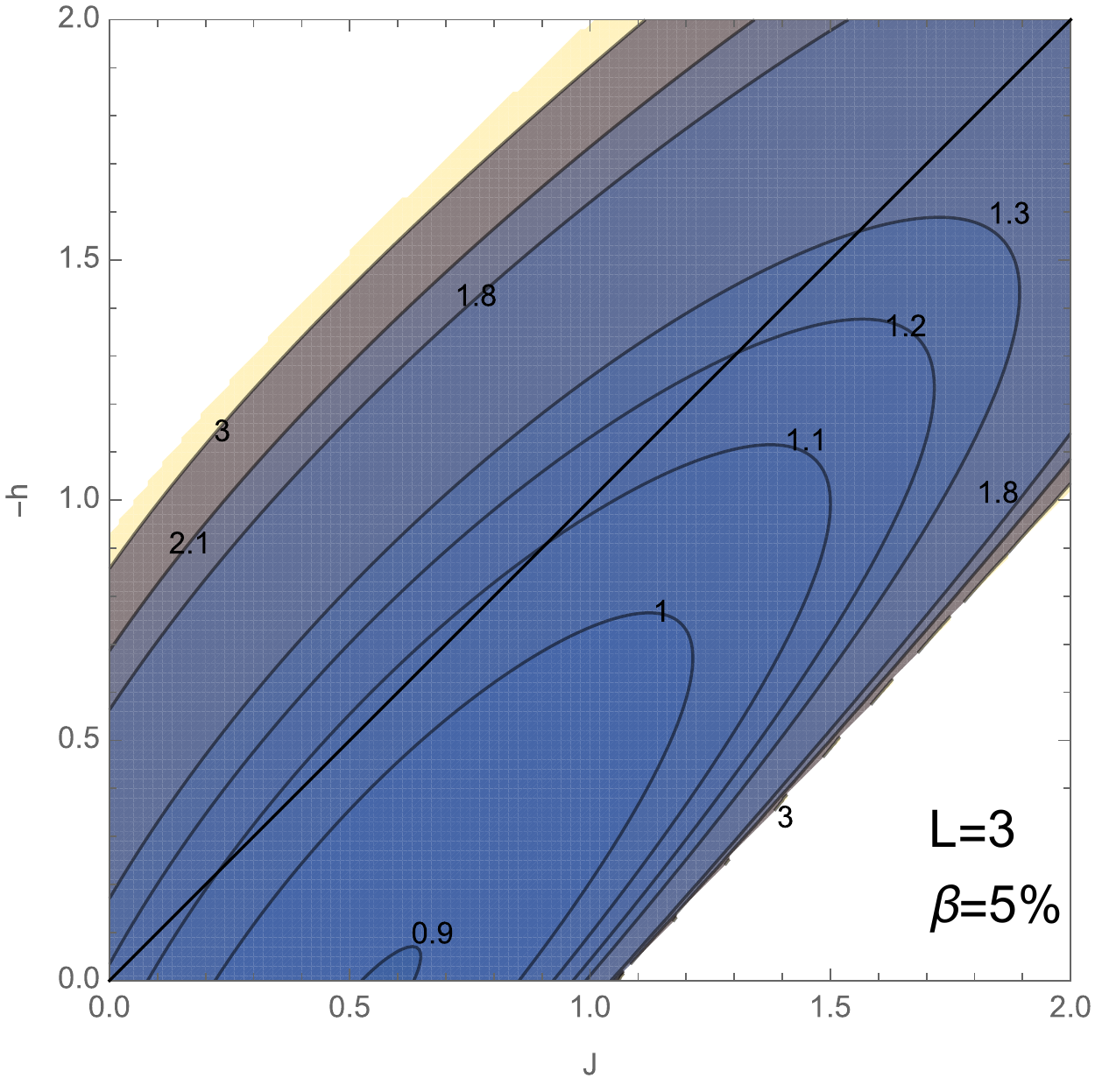}
\includegraphics[width=0.45\columnwidth]{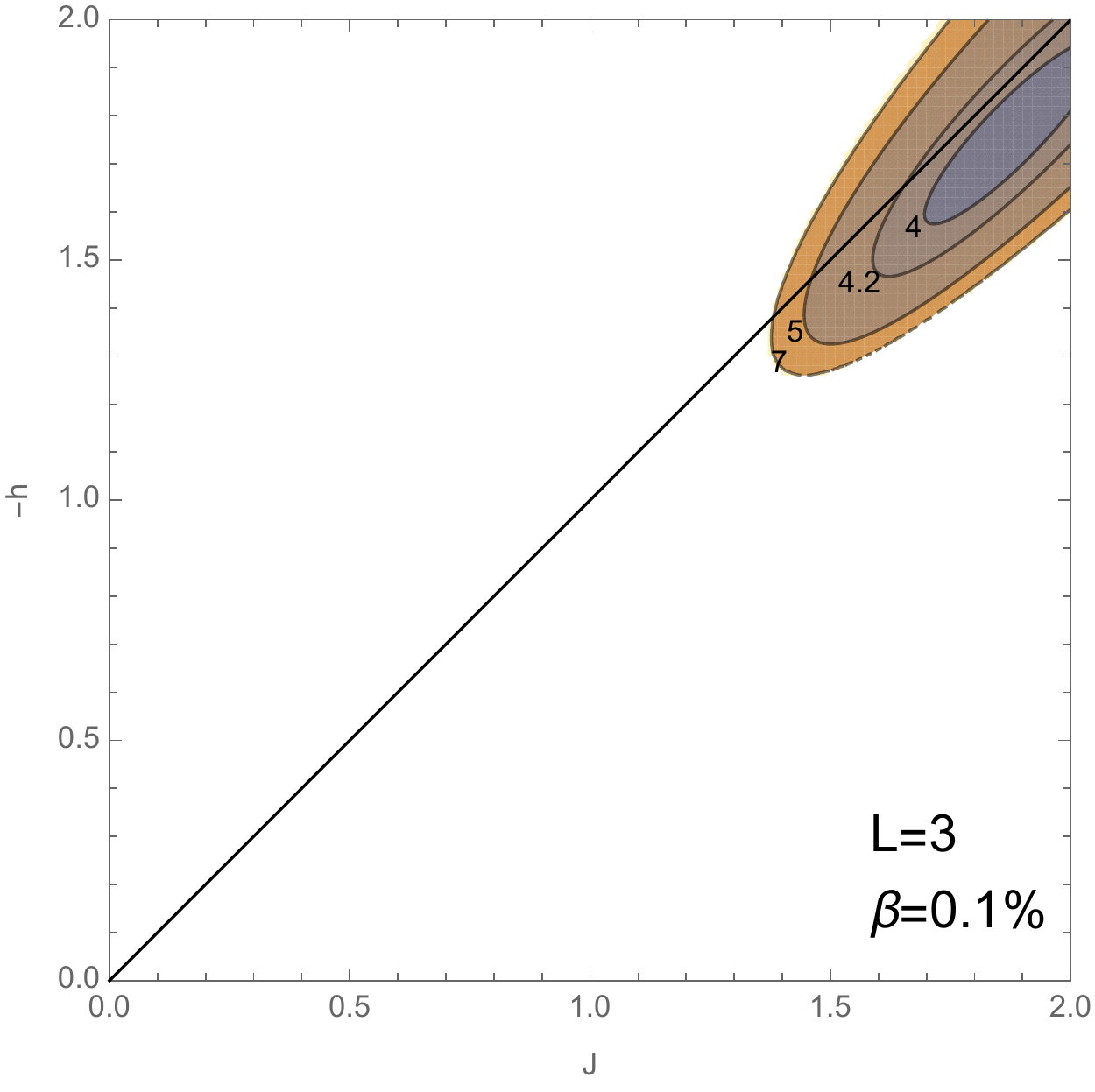}

\includegraphics[width=0.45\columnwidth]{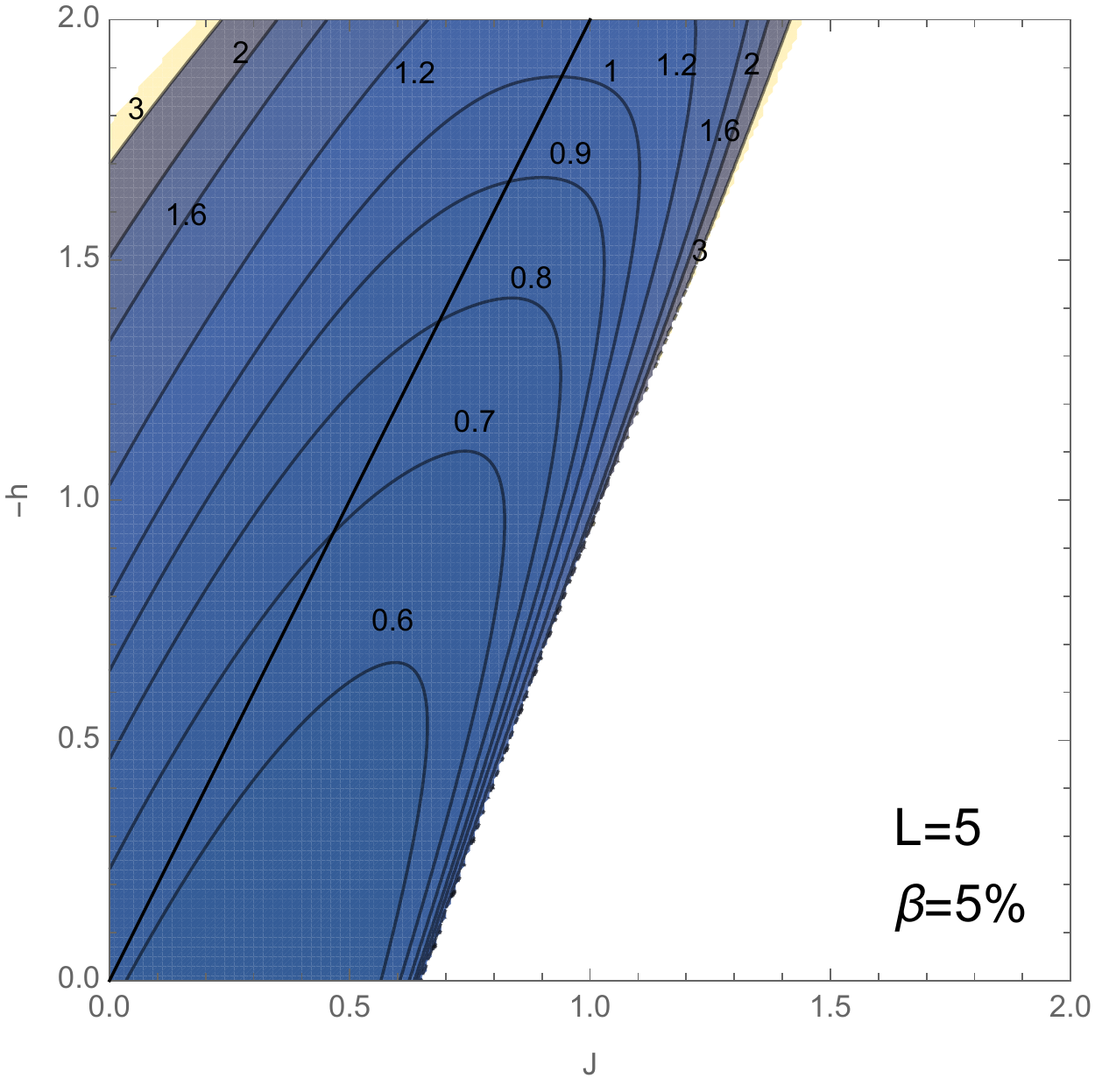}
\includegraphics[width=0.45\columnwidth]{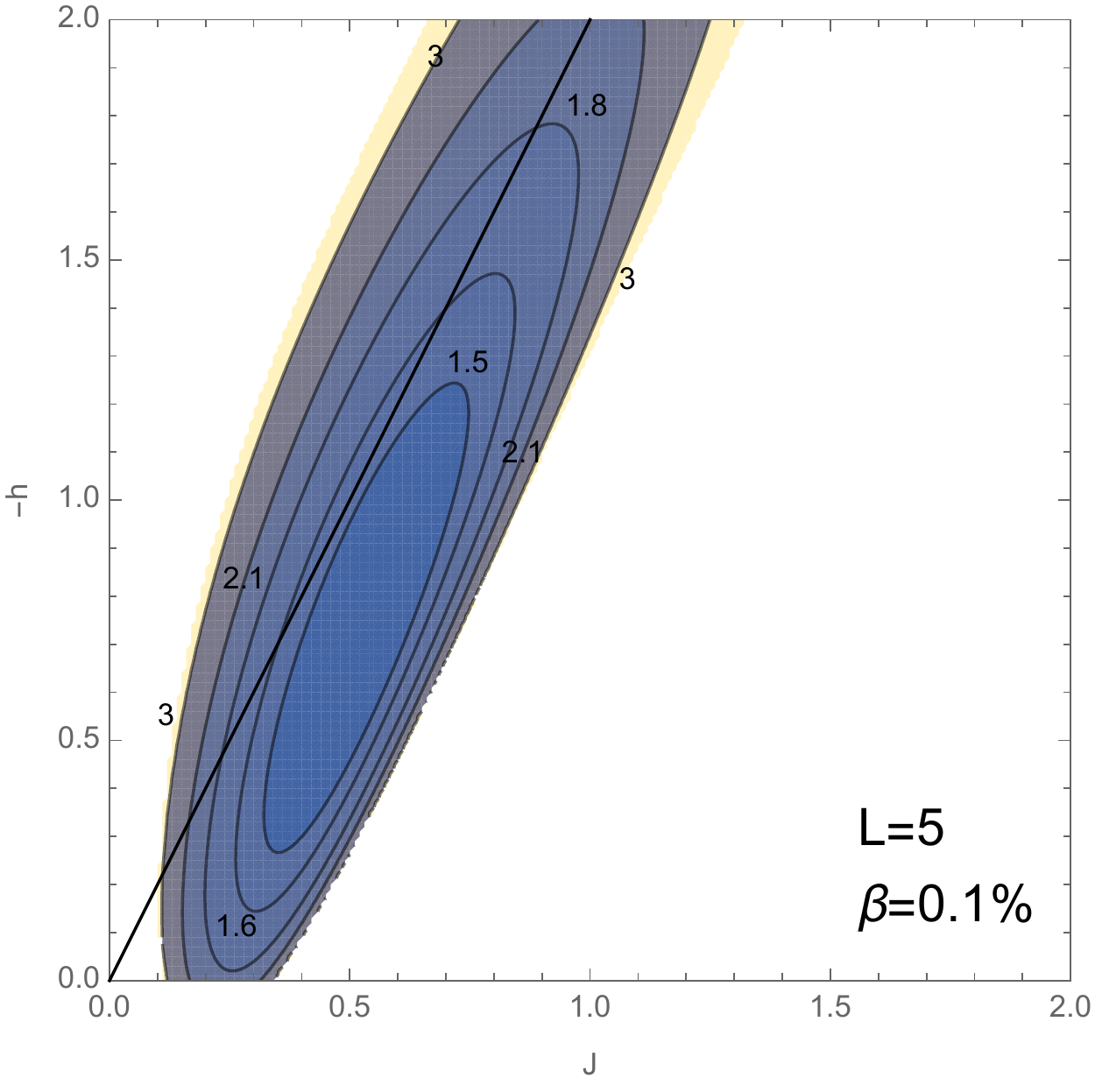}
\caption{Time needed to reach a certain sensitivity as
a function of the coupling coefficient $J$ and minus the external field $-h=\log\frac{K_d}{c}$.
 The solid line indicates  $-h=J(L-1)/2$ which corresponds to the values of the concentration set at the transition point that were used to draw figure~\ref{fig:time_det}.
 The
top figures refer to the case with $L=3$ and the bottom ones to $L=5$. For each value the concentration change to be detected is such that $c'/c=0.9$ corresponding to $\delta h=-0.1$.
Time is expressed in units of $k_u^{-1}$.}\label{fig:timekd}
\end{figure}
 This is the value for which
the system is most sensitive to concentration changes. In order to extend our reasonings to cases in which the system is not optimally tuned, let us 
consider how the detection time depends on the reference concentration (or, equivalently, on the dissociation constant). The results are reported in fig. \ref{fig:timekd}.
For each level of concentration it is possible to determine the range of cooperativity that allows for a given sensitivity and the 
coupling intensity $J$ that reaches it in the shortest time.
The fastest detections are achieved for combinations of concentration and coupling such that the system is close to the transition point but higher concentrations 
are favored. 
This reflects the fact that the concentration change has to be within the dynamic range of the receptor and that higher concentrations imply a faster ligand binding  consequently 
accelerating the dynamics.

\subsection{Non-homogeneous  coupling strength and ligand affinity across the receptors.}
 
In the previous sections we have considered the case in which each receptor has the same coupling strength and ligand affinity. 
In general, it is possible that  different receptors have different $J$ and $K_d$.
Let us consider the case in which the  $J$ and $K_d$ of each receptor are drawn from a distribution $\rho(J,K_{d})$.
If the number of receptors $N$ is large enough so that  each value of the pair $J, K_d$ is sampled several times we can still exploit Stein's lemma for
each receptors subpopulation. The overall miss probability is then an average over the different receptors subpopulations reading
\begin{equation}
%\lim_{N\to\infty} 
\log\beta\sim-N\int \rho(J,K_{d})\, D_{KL}\left(p_{eq}^c(J, K_d)||p_{t}^{c'}(J, K_d) \right)dJ\,dK_{d}\,.
\end{equation}
As a specific example, consider the case where the receptors are  a mixture of  two different subpopulations characterized by two different couplings $J_{w}$ and $J_{s}$ with the
respective $K_d$ tuned as
$K_d(J)=K_d^{(0)}e^{\frac{J}{2}(L-1)}$ so that the transition point is for both populations at $c_{1/2}=K_d^{(0)}$.
%If we keep $k_u$ constant we can use the expressions in eq.~(\ref{eq:k_p}) and (\ref{eq:k_-}) and see that
%\begin{eqnarray}
 %k(l\to l+1)&=&(L-l)k_u \frac{c}{c_{1/2}(J,L)} e^{\frac{  J}{2} \left(l-\frac{L-1}{2}\right)}\\
 %k(l\to l-1)&=&l\,k_u  e^{-\frac{  J}{2} \left(l-\frac{L+1}{2}\right)}
%\end{eqnarray}
\begin{figure}
\includegraphics[width=\columnwidth]{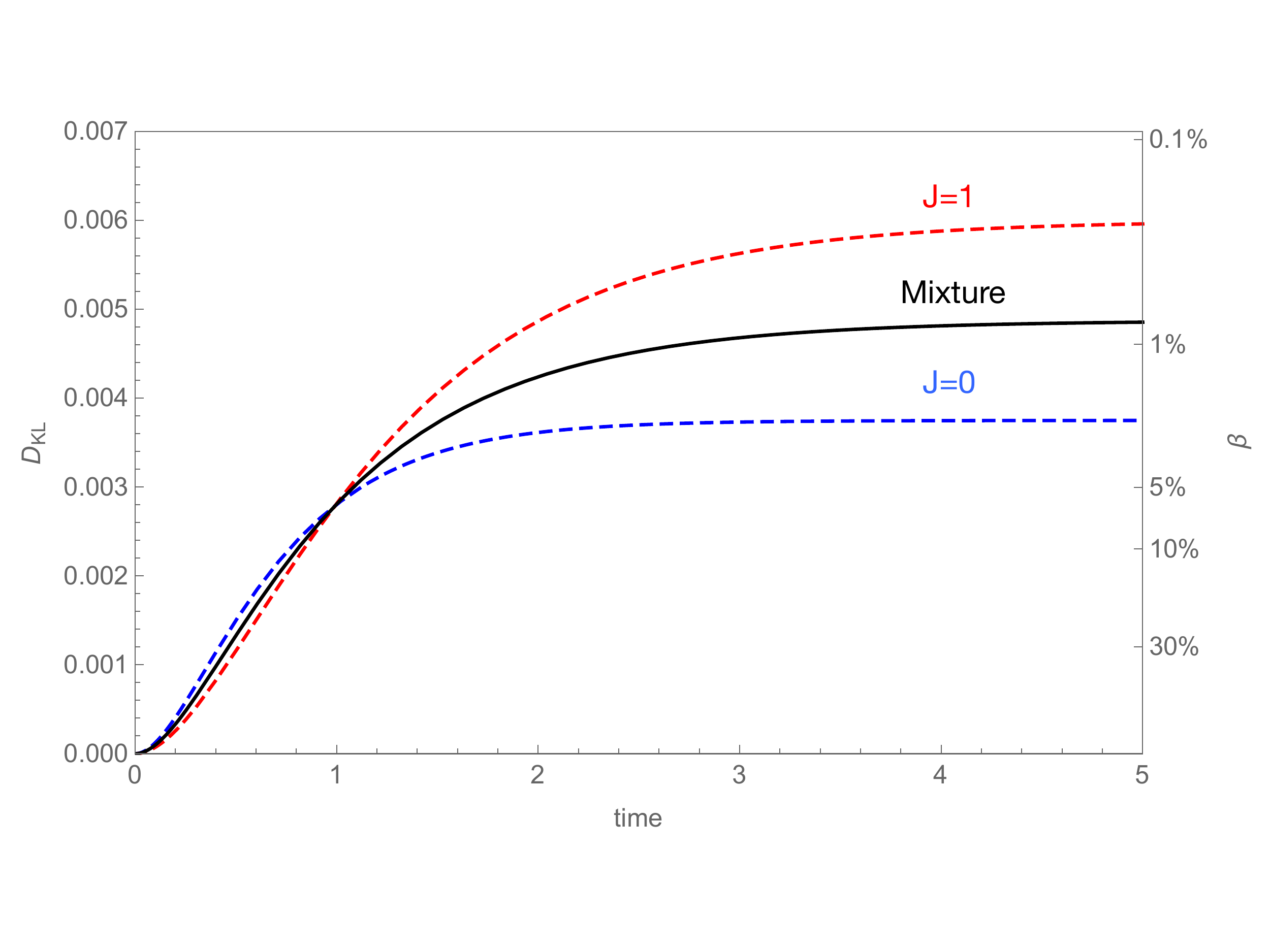}
\caption{In the case of non-homogeneous  coupling strength across the receptors the miss probability is given by a linear combination of the 
various couplings. The plot refers to the case in which a receptor is cooperative with $J=1$ with probability $q=1/2$ and is noncooperative
with probability $1-q=1/2$ with $L=3$ and $c'=c\,e^{-0.1}$. Time is expressed in units of $\left(k_u\sqrt{c'/c_{1/2}}\right)^{-1}$.}\label{fig:mix}
\end{figure}

Let us denote the probability of drawing the strongly cooperative ($J_{s}$) receptor as $q$ so that we have, on average $Nq$ strongly cooperative receptors.
The miss probability then obeys
\begin{equation}
%\lim_{N\to\infty}
 \log \beta\sim-N\left[q D_{KL}\left(p_{eq}^c(J_{s})||p_{t}^{c'}(J_{s}) \right)+(1-q)D_{KL}\left(p_{eq}^c(J_{w})||p_{t}^{c'}(J_{w}) \right) \right]
\end{equation}
resulting in a linear combination of the two Kullback-Leibler divergences. By varying the parameter $q$ from $0$ to $1$ the curve smoothly varies
between the weakly and the strongly cooperative cases. 
In general, such mixture of receptors displays a faster
initial response compared to the strongly cooperative case and a higher asymptotic precision with respect to the weakly cooperative case (see fig~\ref{fig:mix}). 
This allows for intermediate behaviors which can be tuned  to satisfy different requirements by changing the sizes of the subpopulations. 

}

\section{Conclusions and discussion}
We have studied the problem of detecting a change in concentration of a ligand by means of a pool of receptors.
To investigate the detection performance we have addressed the problem in terms of hypothesis testing making use of
 the instantaneous occupancy. We have focused on the probability of missing the detection 
of a change that actually occurred and employed it as a measure of the quality of the sensing system. 
With this formalism we have considered the influence of cooperativity on the sensing performance.
We have identified a twofold effect of cooperativity:
a markedly improved sensitivity in the long time limit and a significant slowing down of the receptors dynamics.
% in agreement with the findings  of Ref.~\cite{skoge11}.  
This is consistent with the findings of Ref.~\cite{skoge11} in which  the sensing performance of locally cooperative receptors
was considered in terms of the precision with which they are able to determine a concentration change by time averaging their history.
We stress here that both the higher asymptotic sensitivity and the slowing down are separately relevant.
Most notably, cooperativity enables to detect changes with sensitivities
that are not attainable by noncooperative receptors. 
When considering the time from the concentration change needed before the instantaneous receptor occupancy
state can provide a given detection sensitivity, the two effects of cooperativity trade off.
% The authors of \cite{skoge11} considered the sensing performance of locally cooperative receptors
% in terms of the precision with which they are able to determine a concentration change by time averaging their history.
In the framework adopted by the authors of \cite{skoge11}, the  effect of the slowdown due to cooperativity on 
the time averaging procedure dominates
the additional sensitivity and the authors concluded that noncooperative receptors perform better than cooperative ones.
On the contrary, we have seen that, casting the detection problem in terms of hypothesis testing by means of the instantaneous
occupancy of a pool of receptors, cooperativity can be beneficial.
We have found that there is an optimal degree of cooperativity which depends on the 
required sensitivity level and on the number of binding sites present in each receptor.
At given number of binding sites we can identify three general 
scenarios depending on the required detection power.
For hard detections the system must be cooperative and the desired sensitivity is reached earlier at finite but 
%To attain a high sensitivity the system must be cooperative and the desired level is reached earlier at finite but
large levels of cooperativity.
For intermediate cases the system can achieve the needed statistical power by {   noncooperative}
binding sites but moderately cooperative receptors are faster at approaching it.   
Easy detections are performed more rapidly by noncooperative receptors. 

% 
% It is not clear whether cells sense by time averaging the receptor history or restrict to the instantaneous state
% so that both 
% It is fascinating to speculate on
% how the progressive appea evolution
% evolution adding a binding site at the time.\\
It is interesting to consider how the cell's detection ability scales with the numbers of receptors and binding sites.
For {   noncooperative} binding sites the probability of missing the detection decreases exponentially with the 
total number of binding sites present in all the receptors $NL$.
We have shown that for cooperative receptors in the long-time limit  
the exponential decrease is more marked and in the infinite coupling limit for small
concentration changes depends quadratically on the number of binding sites present in each 
receptor: $NL^2$.
Since it is costly to produce proteins (see Refs.\cite{dekel05,govern14pnas}) sensing mechanisms that 
improve the performances without increasing the number of required components may represent an advantage.
The different scaling in $N$ and $L$ can then impact on how to invest the limited resources favoring either 
solutions involving few receptors with many binding sites or the opposite. The characterisation of the ensuing trade-offs 
is an issue that surely deserves further study in the near future.

\section*{Acknowledgements}
S.B. acknowledges ICTP {   and the Physics Department and INFN of the University of Turin} for hospitality.

\end{document}